\documentclass[acmsmall]{acmart}

\AtBeginDocument{%
  \providecommand\BibTeX{{%
    \normalfont B\kern-0.5em{\scshape i\kern-0.25em b}\kern-0.8em\TeX}}}

\setcopyright{acmcopyright}
\copyrightyear{2021}
\acmYear{2021}
\acmDOI{10.1145/1122445.1122456}

\usepackage{xcolor}
\usepackage{amsmath,amsthm,float,caption,mathtools}
\usepackage{balance}
\usepackage{graphicx}
\usepackage[colorinlistoftodos]{todonotes}
\usepackage[most]{tcolorbox}
\usepackage[makeroom]{cancel}
\usepackage{algorithm}
\usepackage[noend]{algpseudocode}
\tikzset{main node/.style={circle,fill=blue!20,draw,minimum size=1cm,inner sep=0pt},}
\newcommand{\E}{\mathbb{E}}

\usepackage{subfig}

\newtheorem{theorem}{Theorem}
\newtheorem*{theorem*}{Theorem}
\newtheorem{lemma}{Lemma}
\newtheorem{definition}{Definition}
\newtheorem*{definition*}{Definition}
\newtheorem*{lemma*}{Lemma}

\newtheorem*{corollary*}{Corollary}
\newtheorem{claim}{Claim}
\newtheorem*{claim*}{Claim}

\newtheorem*{condition*}{Condition}
\allowdisplaybreaks

\acmJournal{JACM}
\acmVolume{37}
\acmNumber{4}
\acmArticle{111}
\acmMonth{8}

\begin{document}

\title{Quarantines as a Targeted Immunization Strategy}
%\author{Anonymous Author(s)}
%
% The "author" command and its associated commands are used to define
% the authors and their affiliations.
% Of note is the shared affiliation of the first two authors, and the
% "authornote" and "authornotemark" commands
% used to denote shared contribution to the research.
 \author{Jessica Hoffmann}
 \authornote{Both authors contributed equally to this research.}
 \email{hoffmann@cs.utexas.edu}
 \orcid{}
 \author{Matt Jordan}
 \authornotemark[1]
 \email{mjordan@cs.utexas.edu}
 \affiliation{%
   \institution{The University of Texas at Austin}
 %  \streetaddress{P.O. Box 1212}
 }

 \author{Constantine Caramanis}
 \affiliation{%
   \institution{The University of Texas at Austin}}
 %  \streetaddress{1 Th{\o}rv{\"a}ld Circle}
 \email{constantine@utexas.edu}

%
% By default, the full list of authors will be used in the page
% headers. Often, this list is too long, and will overlap
% other information printed in the page headers. This command allows
% the author to define a more concise list
% of authors' names for this purpose.
\renewcommand{\shortauthors}{Hoffmann and Jordan, et al.}

%
% The abstract is a short summary of the work to be presented in the
% article.
\begin{abstract}
In the context of the recent COVID-19 outbreak, quarantine has been used to "flatten the curve" and slow the spread of the disease. In this paper, we show that this is not the only benefit of quarantine for the mitigation of an SIR epidemic spreading on a graph. Indeed, human contact networks exhibit a powerlaw structure, which means immunizing nodes at random is extremely ineffective at slowing the epidemic, while immunizing high-degree nodes can efficiently guarantee herd immunity.  We theoretically prove that if quarantines are declared at the right moment, high-degree nodes are disproportionately in the Removed state, which is a form of targeted immunization. Even if quarantines are declared too early, subsequent waves of infection spread slower than the first waves. This leads us to propose an opening and closing strategy aiming at immunizing the graph while infecting the minimum number of individuals, guaranteeing the population is now robust to future infections. To the best of our knowledge, this is the only strategy that guarantees herd immunity without requiring vaccines. We extensively verify our results on simulated and real-life networks.
\end{abstract}

%\begin{CCSXML}
%<ccs2012>
%  <concept_id>10010520.10010553.10010562</concept_id>
%  <concept_desc>Computer systems organization~Embedded systems</concept_desc>
%  <concept_significance>500</concept_significance>
%  <concept_id>10010520.10010575.10010755</concept_id>
%  <concept_desc>Computer systems organization~Redundancy</concept_desc>
%  <concept_significance>300</concept_significance>
%  <concept_id>10010520.10010553.10010554</concept_id>
%  <concept_desc>Computer systems organization~Robotics</concept_desc>
%  <concept_significance>100</concept_significance>
%  <concept_id>10003033.10003083.10003095</concept_id>
%  <concept_desc>Networks~Network reliability</concept_desc>
%  <concept_significance>100</concept_significance>
%</ccs2012>
%\end{CCSXML}
%
%\ccsdesc[500]{Computer systems organization~Embedded systems}


\maketitle

\section{Introduction}
Most real-life networks, from the technological (the Internet \cite{chen2002origin,kleinberg2002small}, train routes \cite{sen2003small}, electronic circuits \cite{i2001topology}) to the biological (neural networks \cite{watts1998collective,white1986structure}, protein interaction networks \cite{jeong2001lethality}) exhibit a powerlaw structure. Human networks are no exception (the film actors network \cite{amaral2000classes,watts1998collective}, the telephone-call graph \cite{aiello2000random,aiello2002random}, the sexual contact graph \cite{liljeros2003sexual,liljeros2001web}). Recent tracking studies have confirmed that the network of proximity contact follows the same distribution \cite{Salathe22020,Karamouzas2014}.

When it comes to epidemics, this has terrible consequences. In particular, this means that the epidemic threshold is vanishing \cite{newman2018networks}, which implies that measures to reduce the probability of infection (hand-washing, social distancing, etc.) can slow the spread of the disease, but not stop the outbreak. It also implies that an outbreak is likely to start anew even from a single infected individual ($e.g$. a traveler). Only measures that break the structure of the graph (such as quarantine, closing restaurants and parks, prohibiting groups of more than 10 etc.) can stop the spread. Moreover, powerlaw graphs are incredibly resistant to $random$ failures \cite{albert2000error}, which means that removing individuals at random (for instance if they received a vaccine, or if they caught the disease and developed immunity) will not change these properties. As most countries affected by COVID-19 are relaxing quarantine while many infected individuals remain infected, one might think that the second wave will be identical to the first wave.

However, if powerlaw graphs are resistant to random failure, they are very susceptible to targeted immunization \cite{albert2000error}. As shown in the previously mentioned paper,  vaccinating only a small percentage of the nodes of the graph (the nodes of highest degree) is enough to achieve herd immunity, while it is never achievable with random immunization. The key observation here is that epidemic spread is $not$ equivalent to random immunization. In particular, nodes of high degree get infected much more quickly than nodes of low degree \cite{newman2018networks}. If we let the epidemic run its course, this superspreader phenomenon would just help the epidemic spread rapidly to all the nodes in the network. However, if the epidemic is temporarily halted, say because of a mandatory quarantine, in the SIR model this implies that nodes of high degree will become immunized without having had the possibility of spreading the infection. Multiple quarantines can then become an ersatz to achieving herd immunity (or at least slow its spread and "flatten the curve") before a vaccine is developed.

Note that if we let the disease spread and infect the entire population, we would technically achieve herd immunity. Our goal is then to balance how many people get infected while guaranteeing the population gets immunized. Leveraging results from random graphs analysis and percolation theory, we theoretically show that even a $single$ well-timed quarantine can be used to achieve herd immunity and significantly reduce the fraction of the infected population.

\subsection{Relevant work}
The study of epidemics on graphs is an active field of research. 
An important body of work assumes the underlying graph is known, and focuses on modeling epidemics \cite{DelVicario2016, Wu2018, Gomez-Rodriguez2013, Cheng2014, Zhao2015, liu2019ct}, detecting whether there is an epidemic \cite{Arias-castro2011, Arias-castro, Milling2015, Milling2012, Meirom2014, Leskovec2007, Khim2017}, finding communities \cite{prokhorenkova2019learning, xie2019meta}, localizing the source of the spread \cite{Shah2010, shah2012rumor, shah2010detecting, spencer2015impossibility, wang2014rumor, sridhar2019sequential, dong2019multiple} or instead obfuscating it \cite{fanti2016rumor, Fanti2014, Fanti2017}, or controlling their spread \cite{kolli2019influence, Drakopoulos2014, Drakopoulos2015, Hoffmann2018, Farajtabar2017, wang2019analysis, yan2019conquering, ou2019screen}. The inverse problem, recovering the network from epidemic data, has also been extensively studied \cite{Netrapalli2012,Abrahao2013,Daneshmand2014,pasdeloup2017characterization,Khim2018,hoffmann2019learning,hoffmann2020}.

This paper is interested in immunization strategies. \cite{albert2000error} already showed that immunizing about $3\%$ of the nodes of highest degree is enough to reach herd immunity for any powerlaw graph, no matter the exponent. \cite{cohen2003efficient} proposed a local strategy in case the global structure of the graph is unknown: we pick a fraction $\rho$ of the population; each of those individuals then names someone they personally know. It turns out that the person selected has a higher chance of being well-connected than the person who nominated them, following the well-known principle stating that "your friends have more friends than you." This paper relies on a similar idea: someone infecting someone else is similar to someone nominating someone else, which explains why nodes of highest degree get infected faster. Other immunization strategies have been proposed \cite{holme2004efficient,chen2008finding,schneider2012inverse,buono2015immunization}.

Many scientists have joined the fight against COVID-19, and work has been done to predict the evolution of the epidemic \cite{Fox2020,du2020risk,Sills145,shoeibi2020automated}, predict diagnostics \cite{ahsan2020study,soares2020automatic}, analyze the spread of misinformation about the pandemic online \cite{alam2020fighting,nakov2020spot}, design tracking algorithms and technologies \cite{martin2020demystifying,baumgrtner2020mind,chan2020pact}, analyze testing and interventions policies \cite{lorch2020spatiotemporal,Duque202009033,Duque2020relax,mastakouri2020causal}, and more. One other work \cite{feng2020scheduling} has analyzed quarantine strategies, albeit without taking into consideration the impact on the graph, which is our main contribution. This work aims to analyze quarantines and emphasize their impact as a targeted immunization strategy, which would be robust to future reinfections (contrary to contact tracing strategies, for instance).

\subsection{Main contributions}
The contributions of this paper are as follows:
\begin{itemize}
	\item While quarantines have been introduced to slow the spread of COVID-19, we show that they transform the structure of the human contact graph, and negatively impact the diffusion of the disease during the subsequent waves.
	\item We characterize when to declare the quarantine in order to achieve herd immunity. For simple powerlaw graph of exponent 3 in the configuration model, this corresponds to when about $8\%$ of the graph is infected.
	%\item We experimentally show that it is possible to minimize the expected number of removed individuals at the end of the process. While about 100\% of the nodes contract the disease with commonly accepted COVID-19 parameters, about 65\% of the population will have contracted the disease with a well-timed quarantine. 
	\item While the number above is higher than what we would hope, it is important to note that after this single quarantine, outbreaks cannot start again from a constant number of individuals. As such, this is the first long-term immunization solution which is robust to reinfections but does not require a vaccine.
	\item If we have some limitations on the maximum number of infected individuals ($e.g.$ a limited number of hospital beds), we experimentally show that we can declare multiple quarantines and recover the result from a single quarantine.
	\item We experimentally confirm our results on a wide range of simulated and real-life networks. While our theoretical results are proven on a specific random graph model, we experimentally show that relaxing the theoretical assumptions we require in our proofs (infinite graph, vanishing clustering coefficient, etc.) does not qualitatively impact the results. In particular, well-timed quarantines are a valid targeted immunization strategy for real-life networks.
\end{itemize}

\section{Preliminaries}
\subsection{Notations}
The epidemic spreads on a (possibly infinite) directed graph $G = (V, E)$, where $V$ is the set of nodes and $E$ the set of edges. If $|V| < \infty$, we write $|V| = N$. When considering a random graph drawn from a specific distribution $\mathcal{D}$, we denote by $P_k$ the random variable representing the number of nodes of degree $k$ in the graph, $p_k$ the fraction of nodes of degree $k$, $<k>$ the average degree given the graph, and $q_k$ the fraction of nodes of excess degree $k$, defined by $q_k = \frac{(k+1)p_k}{<k>}$.  As a shorthand, we write $\E_{\{L_k\}_{k=0}^{\infty}}[k] = \displaystyle\sum_{k=0}^{\infty} L_k\cdot k$. If no sequence $\{L_k\}_{k=0}^{\infty}$ is specified, we assume $L_k = P_k$, and $\E[k] = <k>$ and $\E[k^2]$ represent the average degree and degree squared given the graph. Throughout the paper, we use the following functions:
\begin{definition}[generative function]
    For a given graph, the generative function of the degrees $g_0$ and excess degrees $g_1$ are defined as:
    \begin{align*}
        g_0(z) &=  \E_{\{p_k\}_{k=1}^{\infty}}[z^k] = \displaystyle\sum_{k=1}^{\infty} p_k \cdot z^k, \\
        g_1(z) &=  \E_{\{q_k\}_{k=1}^{\infty}}[z^k] = \displaystyle\sum_{k=1}^{\infty} q_k \cdot z^k = \frac{g_0'(z)}{<k>}.
    \end{align*}
\end{definition}
\subsection{Model}
\textbf{SIR model on graphs:} We consider the spread of a Susceptible $\rightarrow$ Infected $\rightarrow$ Removed (SIR) epidemic on a directed graph $G = (V, E)$, possibly infinite. Each node belongs to one of these three states. Infected nodes can infect the susceptible nodes with which they share an edge, also called their $neighbors$ in the graph. Each infection along an edge is independent of other infections. Infected nodes spontaneously transition to the Removed state after a non-deterministic time. Once in the Removed state, nodes do not interact with the epidemic anymore. \\\\
\textbf{Spreading model:} The results presented in this paper are spreading-mechanism agnostic. Our theoretical results do not rely on restrictions on the spreading process. While our experiments are shown on classical continuous-time SIR spread, we do not believe the spreading process would change the outcome. \\\\
\textbf{Configuration model:} Our theoretical results are established in the configuration model \cite{newman2018networks}, an established setting to study random graphs. In this model, we specify the number of nodes $P_k$ of degree $k$. We assume the sum of all $P_k$ is even. We then assign stubs to nodes, such that the number of nodes with $k$ stubs is exactly $P_k$. Following this, we pick two stubs at random, and connect them. We repeat the process until no stubs are left. We say the resulting graph was drawn from the configuration model with degrees sequence $\{P_k \}_{k=0}^{\infty}$. \\\\
\textbf{Powerlaw graphs:} We emphasize the results on powerlaw graphs. A powerlaw gaph of exponent $\alpha$ is a random graph with degree distribution following the law $p_k \sim \frac{C}{k^\alpha}$.  \\\\
\textbf{Simple powerlaw graphs:} We say a graph is a simple powerlaw graph of exponent $\alpha$ if its distribution follows exactly $p_k = \frac{1}{\zeta(\alpha)} \frac{1}{k^\alpha}$ for $k \geq 1, \alpha > 1$. \\\\
\textbf{Barabási–Albert (BA) graph:} A Barabási–Albert (BA) graph of parameter $m$ is constructed by adding nodes one by one, each new node attaching $m$ new edges to previous nodes, picking the node to attach to at random with probability proportional to the degree of the previous nodes (preferential attachment). These graphs follow the configuration model by construction. In expectation, these graphs have powerlaw distribution with exponent 3, with exact degree distribution given by $p_k = \frac{2m(m+1)}{k(k+1)(k+2)} \sim \frac{C}{k^3}$ for $k \geq m$.

\subsection{Known results in the configuration model}\label{sec:prelem}
We recall known results in the configuration model, which we use in the rest of our proofs.

\begin{claim}[Exponential growth, proved in \cite{newman2018networks}]
	In the configuration model, the reproductive number $R = \frac{c_2}{c_1}$, which represents the number of $2^{nd}$ neighbors divided by the number of $1^{st}$ neighbors, can be computed from the sequence of degrees  $\{P_k\}_{k=0}^{N}$. Its value is:
	
	$$R = \frac{\E[k^2] - \E[k]}{\E[k]}.$$
\end{claim}

\begin{claim}[Herd immunity, proved in \cite{newman2018networks}] \label{cl:herdImmunity}
	In the configuration model, an outbreak is possible if and only if $R > 1$, or equivalently:
	$$ \E[k^2] - 2\E[k] > 0.$$
	If the above equation is not satisfied, we have achieved herd immunity.
\end{claim}

\begin{lemma}[Disparity in infection rate by degree, proved in \cite{newman2018networks}]\label{lem:highDegreeU}
	In the configuration model, let $u(t)$ be the expected fraction of nodes of degree 1 in the susceptible state after time $t$. Then as a first order approximation, the expected fraction of nodes of degree $k$ in the susceptible state is $u(t)^k$.
\end{lemma}
\textbf{Remark:} The lemma above is a approximation. In particular, it takes into account how many nodes of degree $k$ become Removed because they got infected, but not that the number of nodes of degree $k$ decreases as their neighbors gets infected (as their degree becomes lower than $k$) and increases whenever nodes of degree $k+c$ get $c$ neighbors in the Removed state.

\subsection{Known results for powerlaw graphs}
We start by recalling the definition of epidemic thresholds:
\begin{definition}[Epidemic threshold, proved in \cite{newman2018networks}]
	For SIR epidemics on graph, there exists a phase transition. If the parameters of the epidemic are above the epidemic threshold, outbreaks occur; otherwise, the epidemic dies out quickly. The closer we are to the epidemic threshold (from above), the smaller the outbreak is. 
\end{definition}

The crucial result about human networks is that epidemic threshold for infinite powerlaw graphs is 0. 
\begin{lemma}[Vanishing threshold, Proved in \cite{newman2018networks}]
Infinite powerlaw graphs with exponents between 2 and 3.4788 have a vanishing epidemic threshold. 
\end{lemma}
The higher the reproductive number $R$ is compared to the epidemic threshold, the more likely an outbreak occurs, and the bigger its expected size is. When the epidemic threshold is 0, this means that changing the parameters of the spread (for instance by enforcing hand-washing  or masks-wearing) can slow the spread of the disease and reduce the total number of infected people, but not prevent that there will be an outbreak. It also means that a single infected traveler can restart a major outbreak. In practice, real graphs are finite, which implies the epidemic threshold is bounded away from 0. However, the larger the graph, the lower the epidemic threshold.

\begin{claim}
	The generative function of the degrees $g_0$ and excess degrees $g_1$ for an infinite powerlaw graph of parameter $\alpha$ is given by:
	\begin{align*}
	g_0(z) &= \frac{1}{\zeta(\alpha)}  \displaystyle\sum_{k=1}^{\infty}\frac{z^k}{k^\alpha} = \frac{Li_\alpha(z)}{\zeta(\alpha)} \\
	g_1(z) &= \frac{1}{\zeta(\alpha - 1)} \displaystyle\sum_{k=1}^{\infty}\frac{z^{k-1}}{k^{\alpha-1}} = \frac{Li_{\alpha - 1}(z)}{z\cdot\zeta(\alpha - 1)}
	\end{align*}
\end{claim}

\section{One quarantine}
In this section, we prove our main result, which is that for certain types of graphs, it is possible to achieve herd immunity by timing a \textit{single} quarantine adequately.
We therefore study the effect of quarantine timing on the structure of the graph. We are particularly interested in whether it is possible to achieve herd immunity, and how to minimize the total number of people who have been touched by the epidemic. For the remainder of the paper, we study perfect quarantines, as defined below:

\begin{definition}[Quarantine]
	We call perfect quarantine (henceforth just quarantine) the complete halt of the spread of the epidemic. During a quarantine, all Infected nodes transition to the Removed state, and no new nodes become Infected.
\end{definition}

\subsection{The quarantine operator}
In the configuration model, the expected behavior of the spread of epidemics is governed by the sequences of degrees $\{P_k\}_{k=0}^{N-1}$. We start by studying the impact of a quarantine on that sequence of degrees.
\begin{lemma}
	Let $u(t)$ be the fraction of nodes of degree 1 that are susceptible after letting the epidemic spread for time $t$. Suppose we declare a quarantine when $u(t)$ reaches the value $u$. Let $T_Q: \mathbb{N}^N \rightarrow \mathbb{R}^N$ be the operator representing the transformation of the  expected number of susceptible nodes after one iteration of letting the epidemic grow, then declaring a quarantine.   Then:
	
	$$ T_Q(\{P_k\}_{k=0}^{N-1}) = \{P_k \cdot u^k\}_{k=0}^{N-1}.$$
	\proof This is a direct implication of Lemma \ref{lem:highDegreeU} from the Preliminaries.
\end{lemma}

As it turns out, the number of nodes in the Removed state can be computed easily from the above result.

\begin{claim}
	Let $g_0(z)$ be the generative function of the distribution of degrees. The expected fraction of nodes in the Removed state after one quarantine $R_Q$ is:
	
	$$ R_Q = 1 - g_0(u).$$
	
	\proof Let $p_k$ be the fraction of nodes of degrees $k$ in the graph, so that $P_k = N\cdot p_k$. Notice that for $k\geq N$, there exist no nodes of that degree, so $p_k = 0$. The generative function of the distribution of degrees is $\displaystyle g_0(z) = \sum_{j=0}^{\infty} p_j z^j$. 
	\begin{align*}
	R_Q &= \frac{1}{N} \left(N - \sum_{j=0}^{N} T_Q(\{P_k\}_{k=0}^{N})_j \right) \\
	&= \frac{1}{N} \left(N  -  \sum_{j=0}^{N} P_j \cdot u^j \right) \\
	 &= 1 -  \sum_{j=0}^{\infty} p_j \cdot u^j \\
	&= 1 - g_0(u).
	\end{align*}
\end{claim}

\subsection{Achieving herd immunity}
\subsubsection{General graphs}
It is possible to give general results for any graph in the configuration model, which we do in this section. 

\begin{claim} \label{cl:generalQuarantine}
	Let $u$ be the remaining fraction of susceptible nodes of degree 1 when we start the quarantine. We achieve herd immunity for $u$ such that:
	$$ u^2 g_0''(u) - u g_0'(u) \leq 0.$$
	\proof Following Claim \ref{cl:herdImmunity}, we want to find $u$ such that $\E_{T_Q}[k^2] - 2 \E_{T_Q}[k] \leq 0$. This translates to:
	\begin{align*}
	\E_{T_Q}[k^2] - 2 \E_{T_Q}[k] &= \sum_{k=0}^{\infty} P_k \cdot u^{k} k^2 - 2 \sum_{k=0}^{\infty} P_k \cdot u^{k} k \\
	&= \sum_{k=0}^{\infty} P_k \cdot u^{k} k^2 -  \sum_{k=0}^{\infty} P_k \cdot u^{k} k -  \sum_{k=0}^{\infty} P_k \cdot u^{k} k \\
	&= \sum_{k=0}^{\infty} P_k \cdot u^{k} k(k-1)  - \sum_{k=0}^{\infty} P_k \cdot u^{k} k \\
	&= u^2 g_0''(u) - u g_0'(u)\\
	&\leq 0.
	\end{align*}
\end{claim}

However, further analysis requires better knowledge of the network topology. We now focus on specific types of random graphs.

\subsubsection{Simple powerlaw graphs}
Here, we provide our first numerical results for simple powerlaw graphs, and show that well-timed quarantine can guarantee herd immunity while drastically reducing the fraction of Infected nodes.
\begin{claim} \label{cl:u}
	Let $u$ be the remaining fraction of susceptible nodes of degree 1 when we start the quarantine. For an infinite simple powerlaw graph of exponent 3, we achieve herd immunity for $u \leq 0.940599$.
	\proof We want to find $u$ such that:
	\begin{align*}
	\E_{T_Q}[k^2] - 2 \E_{T_Q}[k] 
	&= \sum_{k=1}^{\infty} \frac{C}{k^3} \cdot u^{k} k^2 - 2 \sum_{k=1}^{\infty} \frac{C}{k^3} \cdot u^{k} k \\
	&= C \sum_{k=1}^{\infty} \frac{u^{k}}{k}   - 2  \frac{u^{k}}{k^2} \\
	&= -C \cdot \left(2 Li_2(u) - \log(1-u) \right)\\
	&\leq 0.
	\end{align*}
	Where $Li_2(u)$ is the polylog-2 function. Solving this equation numerically yields $u \leq 0.940599$.
\end{claim}

We now look at the total number of nodes in the Removed state in this case:

\begin{claim} \label{cl:rq}
	For a simple infinite powerlaw graph of exponent 3, if we declare a quarantine when a fraction $u' = 0.940599$ of nodes of degree 1 are still in the susceptible state, then by the end of quarantine, at least $R_Q = 92.2912\%$ of the nodes are still in the susceptible state.
	\proof Remembering that $C = \frac{1}{\zeta(3)}$, where $\zeta(x) = \sum_{k=1}^{\infty} \frac{1}{k^x}$, the number of susceptible nodes is: 
	$$g_0(u') \geq \displaystyle\frac{1}{\zeta(3)}\sum_{k=1}^N \frac{u'^k}{k^3} \approx 0.922912. $$
\end{claim}

Combining Claims \ref{cl:u} and \ref{cl:rq}, the network can therefore achieve herd immunity while infecting a bit less than $8\%$ of the nodes:

\begin{theorem}
	For simple infinite powerlaw graphs of exponent 3, min-degree 1, it is possible to achieve herd immunity by declaring a single quarantine when a bit less than $6\%$ of the nodes of degree 1 are Infected. In this case, less than $8\%$ of the nodes will be Removed at the end of the quarantine.
\end{theorem}

\underline{Remarks:} 
\begin{itemize}
	\item  Far from only "flattening the curve", quarantines can be used to achieve herd immunity.
	\item  The result above is both positive (only $8\%$ of Removed nodes) and disappointing: this is the best achievable, in the sense that nothing else we can do in this optimistic quarantine model can reduce the fraction of Removed nodes at the end. In most cases, $8\%$ of the population being Infected represents millions of people, and is not desirable. In the case of COVID-19, however, the fraction of Infected people has passed this number in a number of countries.
	%\item The results above are valid if the original graph have nodes of degree 1. If it does not, the approximations used in Lemma \ref{lem:highDegreeU} are not valid.
	\item The result above is valid for simple powerlaw graph. The exact percentage varies widely with the type of powerlaw graph, as we see below, and in section \ref{sec:experiments}.
\end{itemize}

\subsubsection{Barabási–Albert (BA) graphs}
BA graphs are a well-known family of random powerlaw graphs. We show below that quarantine-based immunization strategies require a higher number of Removed nodes by the time we achieve herd immunity than simple powerlaw graphs (despite having the same min-degree). The exact value of the gain depends on the sequence of of degrees $\{P_k\}_{k=0}^{N-1}$, but the qualitative behavior remains the same. 

\begin{claim}
Let u be the remaining fraction of susceptible nodes of degree 1 when we start the quarantine. For a Barabási–Albert (BA) of parameter $m =1$, we achieve herd immunity if we declare a quarantine when: 
$$u^2 - 4u +3u\log(1-u) -4\log(1-u) \leq 0.$$
Solving numerically, we obtain $u \leq  0.776621$.
\proof Let $g_0(u) = \displaystyle\sum_{k=m}^{\infty} p_k u^k$ be the probability generative function of a BA graph of parameter $m = 1$. We know $p_k = \frac{2m(m+1)}{k(k+1)(k+2)} = \frac{4}{k(k+1)(k+2)}$. Using Claim \ref{cl:generalQuarantine}, we have:
 \begin{align*}
u^2 g_0''(u) - u g_0'(u) &=  u^2\sum_{k=1}^{\infty} \frac{4(k-1)}{(k+1)(k+2)} u^{k-2} -  u\sum_{k=1}^{\infty} \frac{4}{(k+1)(k+2)} u^{k-1}\\
&=  \sum_{k=1}^{\infty} \frac{4(k-2)}{(k+1)(k+2)} u^k \\
&= 4\frac{u^2 - 4u +3u\log(1-u) -4\log(1-u)}{u^2}.
%&=-2 m \cdot u^m \left[ 3 m \phi(u, 1, m + 1) - 4 m \phi(u, 1, m + 2) \right. \\
%&\quad \left.+ 3 \phi(u, 1, m + 1) - 4 \phi(u, 1, m + 2)\right].
\end{align*}
%where $\phi(z,s,a) := \displaystyle\sum_{k=0}^\infty \frac{z^k}{(k+a)^s}$ is the Hurwitz Lerch Transcent, a generalization of the polylogarithm and the Hurwitz zeta function.

Since we need $u^2 g_0''(u) - u g_0'(u) \leq 0$ according to Claim \ref{cl:generalQuarantine}, this is equivalent to $u^2 - 4u +3u\log(1-u) -4\log(1-u) \leq 0$. Numerically, we obtain $0 \leq u \leq  0.776621$.
\end{claim}

\begin{theorem}
	For a BA graphs of parameter $m=1$, it is possible to achieve herd immunity by declaring a single quarantine when a bit less than $23\%$ of the nodes of degree 1 are Infected. In this case, less than $33\%$ of the nodes will be Removed at the end of the quarantine.
\end{theorem}

\subsubsection{Non-powerlaw graphs}
The fact that quarantines can be used as a targeted immunization strategy is a direct result of the heavy-tail distribution of human networks. Here, we illustrate that if contact networks had a different graph structure, we could not achieve herd immunity using a well-timed quarantine without infecting most of the graph -- if not the entire graph. We prove this for Poisson graphs, which do have a tail, albeit lighter than powerlaw graphs, and $d$-regular graphs, which do not have a tail.

\begin{proposition}
For Poisson graphs of parameter $\lambda$ in the configuration model, i.e. with $p_k = \frac{\lambda^k e^{-\lambda}}{k!}$ for $k\geq 0$, it is possible to achieve herd immunity after one quarantine. 
To compare with BA graphs of parameter $m=1$, we pick $\lambda = 2$, so that both graphs will have same average degree. In this case, a bit more than $63\%$ of nodes will be in the Removed state when herd immunity is achieved. 
\proof 
For Poisson graphs in the configuration model, we have:
\begin{align*}
    g_0(x) &= \sum_{k=0}^{\infty} \frac{\lambda^k e^{-\lambda}}{k!} x^k \\
    &= e^{\lambda x - \lambda}.
\end{align*}
Using Claim \ref{cl:generalQuarantine}, we need to declare a quarantine when the fraction of nodes of degree 1 $u$ is such that:
\begin{align*}
   0 &\geq u^2 g_0''(u) - u g_0'(u) \\
   &= u^2 \lambda^2 e^{\lambda x - \lambda} - u \lambda e^{\lambda x - \lambda} \\
   &= u \lambda (u \lambda - 1) e^{\lambda x - \lambda}.
\end{align*}
The epidemic threshold is therefore achieved after one quarantine declared when $u = \frac{1}{\lambda}$. In this case, the total number of nodes in the Removed state at the end of the infection is, in expectation, $1 - e^{\lambda \frac{1}{\lambda} - \lambda} = 1 - e^{1-\lambda}$. When $\lambda = 2$, we obtain $1 - \frac{1}{e} \approx 63.2121\%$.
\\\qed
\end{proposition}

While the tail of Poisson graphs is not as heavy as powerlaw graph, it is still possible to achieve herd immunity, albeit at the cost of infecting a larger fraction of the population. For random d-regular graphs, however, it is impossible to use quarantines as a targeted immunization strategy:

\begin{proposition}
For random $d$-regular graphs in the configuration model, i.e. with $p_d = 1$ and for $k\neq d, p_k = 0$, it is possible to achieve herd immunity after one quarantine for $d>2$. 
\proof See Appendix.
\end{proposition}

\subsubsection{Summary}
We summarize the results above in Table \ref{table:summary}. We notice that experimental value are always below our theoretical predictions. As expected, on powerlaw graphs, Quarantine/Experiments-Q perform worse than High-degree, but significantly better than Random, although they do not require any vaccines. For graphs with lighter tail however, Random performs better. Using quarantine as a targeted immunization strategy is therefore only possible thanks to the very specific distribution of human networks.

\begin{table}[H]
    \centering 
	\begin{tabular}{ lrrrr} 
		\hline
		Graph & Random & \textbf{Quarantine}  & Experiments-Q    & High-degree  \\
		\hline
		%Simple powerlaw/$\alpha = 2, k_{min} = 1$ &  & & & 3\% \\
		 Simple powerlaw/$\alpha = 3, k_{min} = 1$ & 2\% & 8\% &  2\%&  3\% \\
		 %Simple powerlaw/$\alpha = 4, k_{min} = 1$ & 11.13\% & & & 3\% \\
		 Barabási–Albert/$\alpha = 3, m=1$ & 47\% & 33 \% &22\%&  1\% \\
		 Poisson/$\lambda=2$ &  44\%& 65\% & 42\%& 9\%\\
		 4-regular &64\% & -- &  89\%& 64\%\\
		\hline
	\end{tabular} \caption{We show immunization strategies on multiple network topologies: two powerlaw graphs (heavy-tail), a Poisson graph (lighter tail) and a d-regular graph (no tail). The number represent the number of nodes we need to immunize, either by vaccinating them (for Random and High-degree) or by just letting the epidemic spread (for Quarantine). In the Random strategy, we immunize nodes at random. In the High-degree strategy, we immunize nodes by decreasing values of degree. The Quarantine column represents our theoretical predictions, and Experiments-Q the values observed experimentally.}\label{table:summary}
\end{table}

\subsection{Minimizing the final number of Removed nodes}
Below the epidemic threshold, the expected fraction of the population in the Removed state by the end of the epidemic is 0. However, above the epidemic threshold, not all nodes may become Infected by the end of the outbreak, and outbreaks tend to be of small size just above the epidemic threshold. Instead of timing quarantines to reach herd immunity, we might instead allow small outbreaks after our quarantine, provided that it reduces the final number of nodes in the Removed state. In this section, we study how to minimize this final number of Removed nodes, even if that means no reaching herd immunity.

In general, the expected total number of nodes in the Removed state at the end of an epidemic is in general given by:

\begin{claim}[From Newman \cite{newman2018networks}]\label{cl:S}
	Let $\phi$ be the probability that an Infected node infects its susceptible neighbor, and $g_0$ be the generative function of the degrees of the nodes in the graph, and $g_1$ the generative function of the excess degrees. Then the expected number of Removed nodes at the end of the epidemic $S$ is given by:
	$$ S= 1 - g_0(v),$$
	where $v$ is the solution of the equation:
	$$ v = 1 - \phi + \phi g_1(v).$$
\end{claim}

Notice that if $v=1$, we are below the epidemic threshold, and $S=0$. We can compute the generative functions of the degrees and the excess degrees of the graph after a quarantine has been declared:

\begin{claim}\label{cl:g0Q}
	Let $g_0^Q$ and $g_1^Q$ be the generative functions of the degrees and the excess degrees of the resulting graph after a quarantine has been declared when a fraction $u$ of the nodes of degree 1 remain susceptible. Then:
	\begin{align*}
	g_0^Q(z) &= \frac{g_0(u\cdot z)}{g_0(u)}, \\
	g_1^Q(z) &= \frac{g_1(u\cdot z)}{g_1(u)}.
	\end{align*}
	\proof Let $p_k^Q$ and $q_k^Q$ be the distribution of degrees and excess degrees after a quarantine. Using Claim \ref{lem:highDegreeU}, we know $p_k^Q = \frac{u^k \cdot p_k}{g_0(u)}$. Then:
	\begin{align*}
	g_0^Q(z) &= \sum_{k=0}^{\infty} p_k^Q z^k \\
	&= \frac{1}{g_0(u)}\sum_{k=0}^{\infty} p_k \cdot (u \cdot z)^k \\
	&= \frac{g_0(u\cdot z)}{g_0(u)}.
	\end{align*}
	Similarly, since $g_1(z) = \frac{g_0'(z)}{g_0'(1)}$, we have:
%	\begin{align*}
%	g_1^Q(z) &= \frac{g_0^Q'(z)}{g_0'(1)} \\
%	&= \sum_{k=0}^{\infty} \frac{(k+1)p_{k+1}^Q}{<k>} z^k \\
%	&= \frac{1}{g_0(u)}\sum_{k=0}^{\infty} \frac{(k+1)p_{k+1} u^{k+1}}{<k>} z^k \\
%	&= u\frac{u}{g_0(u)} \cdot \sum_{k=0}^{\infty} q_k \cdot  (u \cdot z)^k \\
%	&= \frac{u\cdot g_1(u\cdot z)}{g_0(u)}.
%	\end{align*}
%	
	\begin{align*}
	g_1^Q(z) &= \frac{g_0^{'Q} (z)}{g_0^{'Q} (1)} \\
	&= \frac{u\frac{g_0'(u\cdot z)}{g_0(u)}}{u\frac{g'_0(u\cdot 1)}{g_0(u)}} \\
	&= \frac{g_0'(u\cdot z)}{g'_0(u)} \\
	&= \frac{g_0'(u\cdot z)}{g'_0(1)} \cdot \frac{g'_0(1)}{g'_0(u)} \\
	&= \frac{g_1(u\cdot z)}{g_1(u)}.
	\end{align*}
\end{claim}

Now, it is possible to express the total number of Removed nodes $R(u)$ as a function of $u$, the fraction of susceptible nodes when we declare the quarantine:
\begin{lemma}
	Suppose we declare a quarantine when a fraction $u$ of nodes of degree 1 remain susceptible, then have an outbreak start again after the quarantine. The total expected fraction of Removed nodes $R(u)$ is given by:
	$$ R(u) = 1 - g_0(u\cdot v),$$
	with $v$ solution of $ v  = 1 - \phi + \phi \frac{ g_1(u\cdot v)}{g_1(u)}$.
	\proof 
	Let $R_{\rm beforeQ}(u)$ be the number of Removed nodes at the end of a quarantine, and $R_{\rm afterQ}(u)$ be the expected number of Removed nodes if an outbreak starts again after the quarantine. The total expected fraction of Removed nodes is given by $R(u) = R_{\rm beforeQ}(u) + R_{\rm afterQ}(u)$.
	We know, according to Claim \ref{cl:rq}, that: 
	$$R_{\rm beforeQ}(u) = 1 - g_0(u).$$ 
	We now position ourselves in the graph composed of only the nodes in the Susceptible state after the first quarantine. Notice that this graph is composed of a fraction $g_0(u)$ of the nodes of the original graph. According to Claims \ref{cl:S} and \ref{cl:g0Q}, we also have:
	$$R_{\rm afterQ}(u) = g_0(u) \cdot (1 - g_0^Q(v)) = g_0(u) \cdot \left( 1 - \frac{g_0(u\cdot v)}{g_0(u)} \right) = g_0(u) - g_0(u\cdot v),$$
	with $v$ solution of $ v = 1 - \phi + \phi g_1^Q(v) = 1 - \phi + \phi \frac{g_1(u\cdot v)}{g_1(u)}$.

	Combining the two results, we get $ R(u) = 1 - g_0(u) + g_0(u)- g_0(u\cdot v) = 1 - g_0(u\cdot v)$, which conclude the proof.
	\\\qed
\end{lemma}

The equations above do not have closed form solutions in general, nor for powerlaw graphs in particular. However, we see in the experiment section below that $R(u)$ always seems to exhibit a V shape. If the expected size of an outbreak after a quarantine is very small, it might be better in term of the total number of Infected nodes to declare a quarantine early: letting the infection spread before a quarantine is declared (i.e. when the epidemic is probably in its exponential growth phase) would infect more nodes than having a small outbreak after the quarantine. Minimizing the total number of Infected nodes $R(u)$ is therefore related, but not equivalent to reaching herd immunity. We study $R(u)$ in more details in the next section.

\section{Experiments} \label{sec:experiments}
The remainder of this paper focuses on empirical validation of our theoretical results on a variety of networks, both synthetically generated and taken from real world data sets. We begin by describing the networks examined and the simulation technique. Then we validate the major premise of this work: that higher degree nodes become Infected much more rapidly than lower degree nodes. Then we qualitatively and quantitatively describe the optimal time at which a single quarantine should be enacted, ablating on a variety of simulation parameters. Finally we discuss what is attainable in simulation when multiple quarantines are allowed. 

\subsection{Networks and Simulation Technique}
\paragraph{Simulation setup:} We first describe the simulation environment. All simulations are run using a continuous-time event-driven algorithm modeling an SIR infection \cite{Miller2019}. The infection parameters are $\beta$ and $\gamma$ corresponding to the infection and recovery parameters respectively.  We initialize an infection by uniformly randomly selecting $\rho$ nodes to be Infected. When a quarantine is enacted, every node in the Infected compartment is moved to the Removed compartment and we reinitialize the infection by randomly selecting $\rho$ susceptible nodes to be Infected after the quarantine. As opposed to the previous section, in which quarantines were declared based on the fraction of Susceptible nodes of degree 1 remaining, here we declare quarantines based on the fraction of the general population that has already been Infected (i.e., in the Infected or Removed state, but not in the Susceptible state). We refer to a "quarantine threshold" as the fraction of the general population not in the Susceptible state when we declare the quarantine. All simulations are run until there are no more Infected nodes. Unless otherwise noted, we examine synthetic graphs with 10K nodes and set $\rho$ to be equal to 10. By default, we set the recovery parameter $\gamma$ to be 1 and primarily focus on the modest regime with infection parameter $\beta$ as $\frac{1}{2}$. This places us in the regime where the infection typically spreads throughout the entire population, though we also ablate against varying ratios of $\frac{\beta}{\gamma}$.

\paragraph{Synthetic Networks:} We run our simulations on a wide variety of synthetic networks. We describe these classes of networks here, briefly describing their structure and construction, and finally discuss a realistic choice for parameters. 

The classical model proposed by Barabasi and Albert produces a graph with a power-law degree distribution with exponent $3$ \cite{barabasi1999emergence}. This model operates via an incremental growth mechanism leveraging preferential attachment: new nodes are added incrementally and connected to a fixed number $m$ of existing nodes, where the probability of connection to a node is proportional to the degree of that node. There are two parameters of interest here: $N$, the number of nodes, and $m$, the number of edges added upon the inclusion of a new node. We refer to such graphs as \textbf{BA graphs}. 

While BA graphs have, by design, scale-free degree distributions that emulate many real-world networks, they typically do not have the clustering behavior of real world networks. The clustering coefficient introduced by Watts and Strogatz quantifies this behavior as follows. The local clustering coefficient of node $i$ in a network is defined as the number of edges contained within the neighborhood of node $i$, divided by the number of edges in a clique of the same size. The global clustering coefficient is the average local clustering coefficient over all nodes in the network. The Watts Strogatz network model generates a random graph with small average path lengths and more realistic clustering coefficients, however does not yield a graph with a scale-free degree distribution \cite{watts1998collective}. We refer to such graphs as \textbf{WS graphs}.

To rectify this situation and yield a network with a scale-free degree distribution and controllable clustering coefficients, Holme and Kim modified the BA network model \cite{holme2002growing}. This graph model, which we shall refer to as the power-law cluster, or \textbf{PLC}, model, follows the incremental growth and preferential attachment steps of the algorithm used to generate BA graphs, with one additional step: upon adding a new edge, say to node $j$, with a specified probability, a new edge is added to a neighbor of $j$. This network model has three parameters: $N$, the number of nodes; $m$, the number of edges added by the preferential attachment step; and $p$, the probability of adding a new edge to a neighbor. The clustering coefficient is controlled by the parameter $p$. 

Similar to the PLC model is the random-walk model, \textbf{RW}, first introduced by V\'azquez, which aims to generate a network in a method that emulates organic edge creation in a social network \cite{vazquez2003growing}. New nodes are added incrementally, where each node selects a random extant node to perform a random walk on. At each step, the random walk terminates with fixed probability, and each node in the random walk is connected to the new node with a fixed probability. This model generates networks with scale-free degree distributions and tunable clustering coefficients. The model has parameters: $N$, specifying the size of the graph; $q_e$, the probability of continuing the random walk; and $q_v$, the probability of adding an edge for each element of the random walk. 

Finally, we consider the Nearest Neighbor model that emulates the behavior that two people with a mutual friend are more likely to become friends \cite{vazquez2003growing}. New nodes are connected to a random extant node and a random selection of 2-hop neighbors of the connected node. We use the modification where we also connect $k$ pairs of random nodes upon addition of each graph. This process yields networks with scale-free degree distribution, and tunable clustering \cite{sala2010measurement}. This model has parameters: $N$

\paragraph{Synthetic Network Parameter Selection:}
Each of the network models described above have at least one tunable parameter in addition to their parameter controlling network size. This begs the question of how these parameters should be chosen to emulate real-world social networks, such that a reasonable epidemic simulation may be performed. Fortunately, there is depth of prior work in fitting random network models to real world networks \cite{sala2010measurement, mccloskey2016reconstructing}. We typically use the parameter as specified in \cite{sala2010measurement} that have been fit to a region-based subnetwork from a private Facebook social graph. These parameters, as well as the average degree, clustering coefficient, and best-fit scale-free parameter are described in Table \ref{table:synth-net-table}.

\begin{table}

\begin{tabular}{llrrrr}
\toprule
 Net Type   & Params                 &   Deg &   Cluster &   Pathlength &   Powerlaw Exp. \\
\midrule
 BA         & ($m=5$)                &  9.99 &     0.007 &         3.66 &            2.94 \\
 BA         & ($m=10$)               & 19.98 &     0.011 &         3.06 &            2.98 \\
 NN         & ($u=0.88, k=6$)        & 26.29 &     0.124 &         3.41 &            2.62 \\
 PLC        & ($m=5, p=0.5$)         &  9.99 &     0.178 &         3.53 &            2.67 \\
 PLC        & ($m=10, p=0.25$)       & 19.96 &     0.059 &         2.97 &            2.76 \\
 RW         & ($q_e=0.91, q_v=0.94$) & 19.32 &     0.285 &         3.45 &            2.76 \\
 WS         & ($k=10, p=0.05$)       & 10    &     0.574 &         7.47 &           12.92 \\
\bottomrule

\end{tabular}
\caption{Summary of Synthetic Networks considered. The parameters used for construction are provided, and summary statistics of the average degree, clustering coefficient, and average length of shortest path between two randomly selected nodes. We also use cumulative binning to estimate the powerlaw exponent for the degree distribution. All networks have 10k nodes. While these networks are randomly constructed, the summary statistics concentrate tightly around the reported values. }
\label{table:synth-net-table}
\vspace{-1em}
\end{table}

\paragraph{Real-world networks:} 
We also perform simulations on three main classes of social networks. First we consider the large social network datasets on Facebook and Deezer, as presented in \cite{rozemberczki2019gemsec}. For Facebook, an edge denotes a mutual like for a verified facebook page of a particular category; for Deezer, edges are friendships. Next, we consider the Arxiv physics collaboration datasets taken from \cite{leskovec2007graph}. The data covers papers on Arxiv from the period of January 1993 and April 2003. Here nodes represent authors and a node exists between two authors if they appear on a paper together. All datasets were collected via SNAP \cite{snapnets}.

\begin{table}
\begin{tabular}{llrrrr}
\toprule
 Net Type      & Nodes   &   Deg &   Cluster &   Pathlength &   Powerlaw Exp. \\
\midrule
 arxiv.AstroPh & 18,770   & 21.11 &     0.631 &         4.19 &            2.83 \\
 arxiv.CondMat & 23,131   &  8.08 &     0.633 &         5.36 &            3.11 \\
 arxiv.GrQc    & 5,240    &  5.53 &     0.53  &         6.05 &            2.88 \\
 arxiv.HepPh   & 12,006   & 19.74 &     0.611 &         4.67 &            2.31 \\
 arxiv.HepTh   & 9,875    &  5.26 &     0.471 &         5.94 &            3.28 \\
 deezer.HR     & 54,573   & 18.26 &     0.136 &         4.5  &            3.39 \\
 deezer.HU     & 47,538   &  9.38 &     0.116 &         5.34 &            3.79 \\
 deezer.RO     & 41,773   &  6.02 &     0.091 &         6.35 &            3.7  \\
 fb.artist     & 50,515   & 32.44 &     0.138 &         3.69 &            2.64 \\
 fb.athletes   & 13,866   & 12.53 &     0.276 &         4.28 &            2.85 \\
 fb.company    & 14,113   &  7.41 &     0.239 &         5.31 &            2.97 \\
 fb.government & 7,057    & 25.35 &     0.411 &         3.78 &            2.58 \\
 fb.new\_sites  & 27,917   & 14.78 &     0.295 &         4.39 &            2.85 \\
\bottomrule
\end{tabular}
\caption{Summary of real networks used. The summary statistics are the same as in Table \ref{table:synth-net-table}.}
\label{table:real-net-table}
\end{table}

\begin{comment}
Finally we consider real epidemiological networks. We have obtained the dataset generated by \textcolor{red}{CITE} where real close proximity interactions between students and teachers at an American high school are collected using wireless sensors. We occasionally select only a subset of edges, as per the original paper, collecting only interactions that occur for a specified duration. We also consider the HIV contact network from the Colorado Springs network. We note that while both of these networks are small (N=788, 250, respectively), there is much work in generating synthetic networks that emulate the properties of small input graphs \textcolor{red}{CITE}. We also use ReCoN to generate larger graphs based on these human contact graphs. A full summary of the real networks used is contained in Table \ref{table:real-net-table}.
\end{comment}

\subsection{Groupwise Survival Rates}
\begin{figure}%
    \centering
    \subfloat{{\includegraphics[width=.45\textwidth]{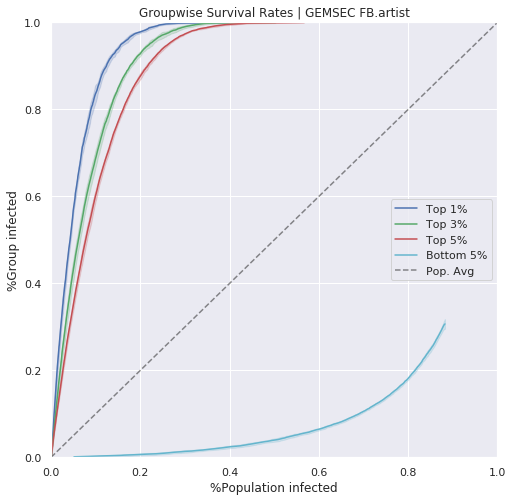} }}%
    \qquad
    \subfloat{{\includegraphics[width=.45\textwidth]{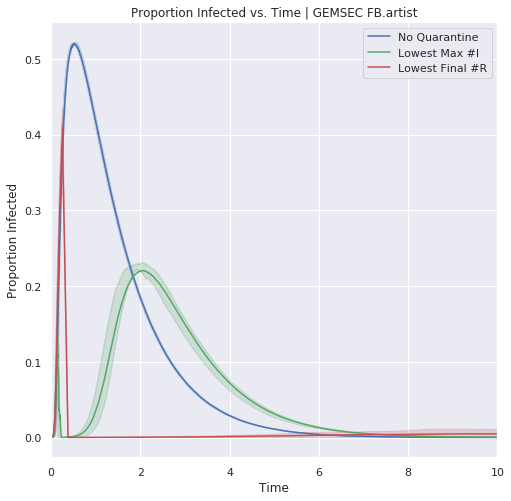} }}
    \caption{Experiments run on the GEMSEC Facebook Artist network. On the left, we see groupwise survival rates. Observe that the higher degree nodes become Infected much more quickly than the population average. On the right, we observe the proportion of simultaneously Infected population versus time. The Blue curves indicate the no quarantine setting; the green curves minimize the peak height, whereas the red curves minimize the total number of Infected nodes.}%
    \label{fig:groupwise-survival-fb}%
\end{figure}

We start with an experiment that motivates the premise of this work: higher degree nodes are more readily Infected, and lower-degree nodes survive much longer. This was shown in \cite{newman2018networks} in the configuration model, but we verify that it still applies even when the configuration model assumptions are violated (e.g. for real-life networks). As mentioned in \ref{sec:prelem}, this behavior is an approximation. We verify the approximation holds for a real-world Facebook graph in Figure \ref{fig:groupwise-survival-fb} (left). In this plot, we have grouped the 1,3,5\% of nodes with highest degree and the nodes with the lowest degree, e.g. 10 for the BA graphwith $(m=10)$. The vertical axis corresponds to the fraction of that group that resides in either the Infected or Removed compartment, and the horizontal axis corresponds to the fraction of the entire population in the Infected or Removed compartments. The dotted line corresponds to the case where the group considered is the entire population. Curves that lie above the dotted line indicate groups that are Infected more rapidly than average nodes. We plot a similar result for a synthetic BA graph with $(m=10)$ in the appendix. We remark that the considered Facebook graph is even starker than the BA graph, suggesting that this effect is stronger in real networks than some of the synthetic networks considered. 

% \begin{figure}%
%     \centering
%     \subfloat{{\includegraphics[width=.45\textwidth ]{figures/vanilla_I_ba10k_10.png} }}%
%         \qquad
%     \subfloat{{\includegraphics[width=.45\textwidth]{figures/vanilla_I_fbartist.png} }}%
%     \caption{Proportion of simultaneously Infected population versus time for a BA network with parameter $m=10$ (left) and the GEMSEC Facebook Artist network (right). The Blue curves indicate the no quarantine setting; the green curves minimize the peak height, whereas the red curves minimize the total number of Infected nodes.}%
%     \label{fig:vanilla-singleQ}%
% \end{figure}

\subsection{Single Quarantine Results} 
We first present results when only a single quarantine was enacted. Throughout we focus our attention on the  GEMSEC Facebook Artist network, a real-life network. Most experiments are also conducted on Barabasi-Albert networks with parameter $m=10$, and can be found on the Appendix. We ultimately demonstrate that the trends exhibited in these networks also are present in the other networks described in Tables \ref{table:synth-net-table} and \ref{table:real-net-table}. 

We outline our single quarantine results. First we demonstrate the curves of the number of Infected nodes versus time to demonstrate the effects of a well-placed quarantine. Then we describe what happens to the total and maximum number of nodes Infected as we vary when the quarantine is enacted, arguing that there are three distinct qualitative quarantine regimes. Next we focus more closely on the structure of a surviving network once an optimally placed quarantine is applied. Then we describe the properties of the second wave of an epidemic that occurs after a quarantine is enacted. Finally we ablate against the infectiousness parameter to demonstrate how a weaker or stronger infection affects the aforementioned results.  

\subsubsection{Survival Versus Time}

We consider the GEMSEC Facebook artist graph. For this network, in Figure \ref{fig:groupwise-survival-fb} (right) we plot the number of nodes currently Infected versus time under three quarantine settings. In blue, we plot the no-quarantine scenario as a baseline. In red we plot the quarantine strategy that minimizes the total number of nodes who become Infected, and in green we plot the quarantine strategy that minimizes the maximum number of nodes who are Infected at any one time. The blue curve describes one potential goal for controlling an epidemic: minimizing the total number of people who ever are Infected by the disease. Whereas the green curve describes another potential goal that gained particular attention in the COVID-19 pandemic: ensuring that the medical system was not overwhelmed at any one time. A similar plot is shown in the Appendix, where we note these curves are qualitatively similar between the real and synthetic networks.

\subsubsection{Varying the Quarantine Threshold}
\begin{figure}%
    \centering
    \subfloat{{\includegraphics[width=.45\textwidth ]{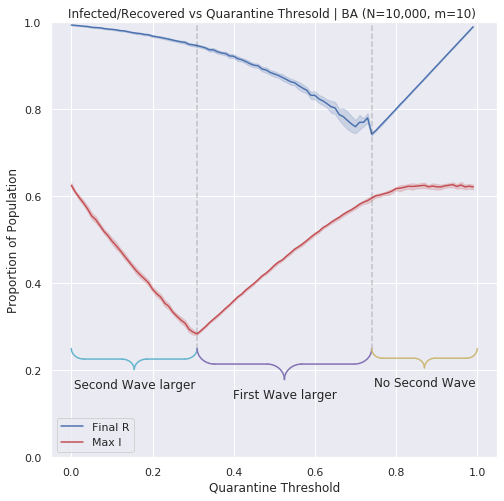} }}%
        \qquad
    \subfloat{{\includegraphics[width=.45\textwidth]{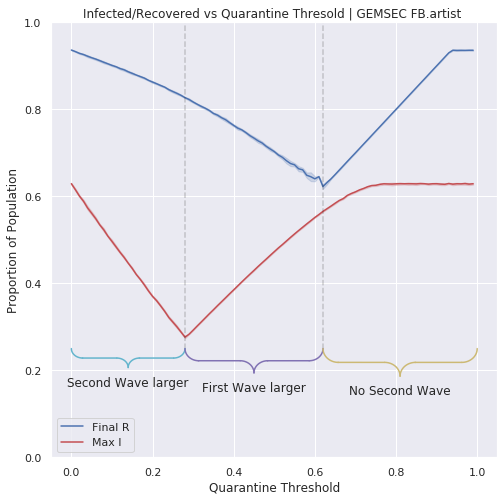} }}%
    \caption{Proportion of population Infected in total (blue curves), and maximum simultaneously Infected population (red curves) versus the quarantine threshold. The left plot is for the BA network, and the right plot is for the GEMSEC Fb.Artist network. Notice that these graphs are qualitatively similar and there are three distinct regions. In the first region, the quarantine is enacted early and the second wave dominates. In the second region, the quarantine is enacted such that the second wave is larger than the first wave, and in the third region there is no second wave. To minimize the number of Infected nodes in total, there should be no second wave.}%
    \label{fig:single-v}%
\end{figure}

Continuing with our primary examples of GEMSEC.fb.artist and BA10 graphs, we plot the total number of nodes who were Infected versus the quarantine threshold in Figure \ref{fig:single-v}. In each plot, the red curve represents the maximum number of simultaneously Infected nodes, whereas the blue curve represents the total number of Infected nodes. We notice three distinct regions along these curves. The first region occurs when the quarantine occurs very early and not enough of the high-degree nodes of the network have been affected. In this case, the second epidemic dominates and the "second wave" has a larger peak. The minimum of the red curve denotes the boundary between the first and second region and corresponds to the situation where the peak-height of the first and second waves are balanced. The second region occurs between the minimum of the red and blue curves and represents the situation where the first wave has a larger peak but the quarantine is effective in reducing the total number of Infected nodes. The optimal number of total Infected nodes is attained at the earliest quarantine that ensures that there is no second wave. This is observable by noticing that the blue curve in the third region is colinear with the identity line: if there were a secondary infection, the blue curve would lie strictly above the identity line. In the right-hand plot of Figure \ref{fig:single-v} we note that there is a flat region of the blue curve when the quarantine threshold is very high: this is because the epidemic parameters are such that the population is never entirely Infected.

\begin{figure}%
    \centering
    \subfloat{{\includegraphics[width=.27\textwidth ]{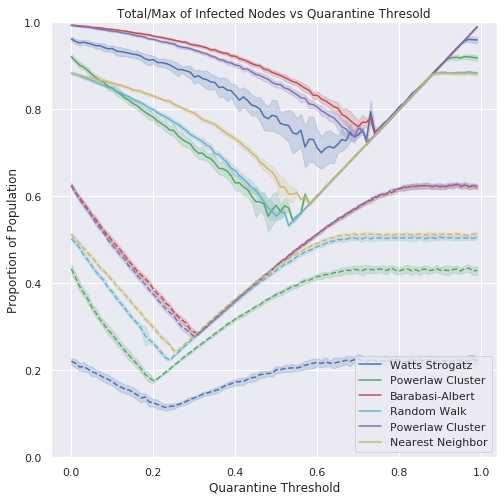} }}%
        \qquad
    \subfloat{{\includegraphics[width=.27\textwidth]{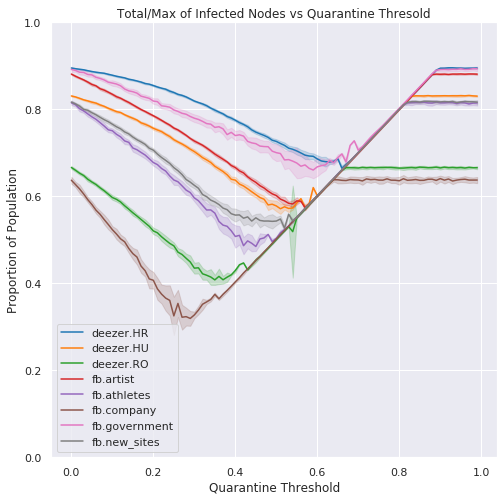} }}%
            \qquad
    \subfloat{{\includegraphics[width=.27\textwidth]{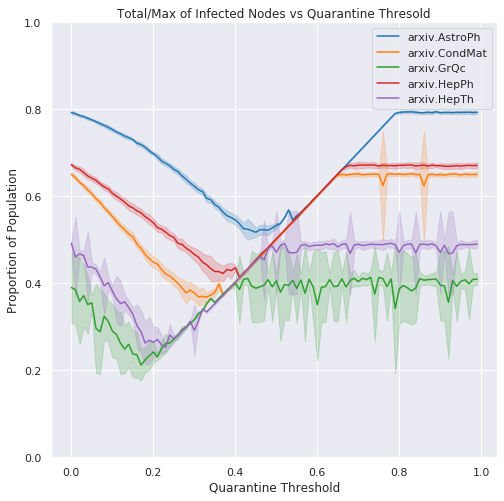} }}%
    \caption{We plot the same graphs as in Figure \ref{fig:single-v}, for the whole suite of synthetic networks (left), the GEMSEC networks (middle) and the Arxiv Collaboration networks. We only plot the total Infected curves (blue in Figure \ref{fig:single-v}). We note that while varying graphs may have various quarantine strategy and efficacy, the qualitative properties remain.}%
    \label{fig:multi-v}%
\end{figure}

To demonstrate that the trends shown above do not apply only to these archetypal networks, we present the total number of Infected nodes versus quarantine threshold for the entire suite of synthetic and real networks in Figure \ref{fig:multi-v}. Notice that each of these networks display the same qualitative features but differ in their height, maximal benefit of quarantine, and location of optimal quarantine threshold.

\subsubsection{Structural Changes After a Quarantine} 
Now we examine the structure of the remaining subgraph after an optimal quarantine has been enacted. In Table \ref{table:synthetic-changes} we compare various graph metrics between the original network and the network that results after performing the single-quarantine that minimizes the total number of Infected nodes. We consider the average degree and average shortest path. Average shortest path is taken by considering 100,000 random pairs of nodes in the largest connected component and computing their shortest path. We notice that for all synthetic graphs, except Watts Strogatz, the average degree reduces dramatically with the optimal-quarantine networks having degrees between 1 and 2. The shortest paths increase by a factor of 3-4x, indicating that these networks become much more tree-like and have a lesser small-world-effect, thus it is no surprise that a second wave cannot occur. In table \ref{table:real-changes} we show similar results for real networks. This indicates that the reductions in average degree and increase in shortest path length under an optimal quarantine are present in a realistic setting as well, albeit to not so drastic an extent.
\begin{table}
\begin{tabular}{llrrlrrl}
\toprule
 Net Type   & Params                 &   Deg &   Deg' & \% Change   &   S.P. &   S.P' & \% Change   \\
\midrule
 BA         & ($m=5$)                &  9.99 &   1.19 & -88.10\%    &   3.68 &  18.58 & +405.39\%   \\
 BA         & ($m=10$)               & 19.98 &   1.3  & -93.51\%    &   3.06 &  14.92 & +388.31\%   \\
 NN         & ($u=0.88, k=6$)        & 26.09 &   1.05 & -95.99\%    &   3.41 &  17.86 & +424.43\%   \\
 PLC        & ($m=5, p=0.5$)         &  9.99 &   1.39 & -86.13\%    &   3.51 &  18.12 & +416.18\%   \\
 PLC        & ($m=10, p=0.25$)       & 19.96 &   1.29 & -93.54\%    &   2.97 &  13.97 & +370.70\%   \\
 RW         & ($q_e=0.91, q_v=0.94$) & 18.95 &   1.65 & -91.29\%    &   3.5  &  17.65 & +404.89\%   \\
 WS         & ($k=10, p=0.05$)       & 10    &   5.09 & -49.07\%    &   7.45 &  11.71 & +57.13\%    \\
\bottomrule
\end{tabular}
\caption{Change of summary statistics between generated graphs and the subgraph of susceptible nodes only after an optimal quarantine was performed (denoted with a superscript tick). As noted, an optimal quarantine ensures that no second wave is possible, which the summary statistics reflect: the networks become much sparser and the small-world effect is dampened.}
\label{table:synthetic-changes}
\end{table}

\begin{table}
\begin{tabular}{lrrlrrl}
\toprule
 Net Type      &   Deg &   Deg' & \% Change   &   S.P. &   S.P' & \% Change   \\
\midrule
 arxiv.AstroPh & 21.11 &   3.09 & -85.37\%    &   4.17 &  10.29 & +146.43\%   \\
 arxiv.CondMat &  8.08 &   2.87 & -64.53\%    &   5.35 &  12.41 & +131.97\%   \\
 arxiv.GrQc    &  5.53 &   3.23 & -41.57\%    &   6.06 &  10.89 & +79.58\%    \\
 arxiv.HepPh   & 19.74 &   2.73 & -86.16\%    &   4.67 &  10.95 & +134.59\%   \\
 arxiv.HepTh   &  5.26 &   2.31 & -56.19\%    &   5.96 &  12.6  & +111.55\%   \\
 deezer.HR     & 18.26 &   2.02 & -88.96\%    &   4.52 &  10.39 & +129.89\%   \\
 deezer.HU     &  9.38 &   2.05 & -78.14\%    &   5.33 &  11.48 & +115.21\%   \\
 deezer.RO     &  6.02 &   1.94 & -67.85\%    &   6.36 &  14.46 & +127.39\%   \\
 fb.artist     & 32.44 &   1.67 & -94.85\%    &   3.69 &  12.97 & +251.16\%   \\
 fb.athletes   & 12.53 &   2.21 & -82.39\%    &   4.28 &  10.4  & +142.71\%   \\
 fb.company    &  7.41 &   2.67 & -63.93\%    &   5.29 &  12.82 & +142.10\%   \\
 fb.government & 25.35 &   2.56 & -89.92\%    &   3.77 &   9.99 & +164.77\%   \\
 fb.new\_sites  & 14.78 &   2.48 & -83.23\%    &   4.39 &  12.93 & +194.31\%   \\
\bottomrule
\end{tabular}
\caption{Change of summary statistics between the real networks and their optimal-quarantine subgraphs. On real networks we see the same qualitative effect as the synthetic networks (c.f.Table \ref{table:synthetic-changes}) but to a lesser extent.}
\label{table:real-changes}
\vspace{-2em}
\end{table}

\subsubsection{Characterizing the secondary infection}
\begin{figure}%
    \centering
    \subfloat{{\includegraphics[width=.45\textwidth ]{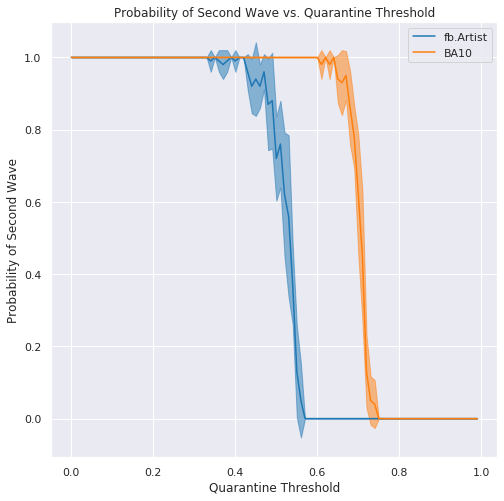} }}%
        \qquad
    \subfloat{{\includegraphics[width=.45\textwidth]{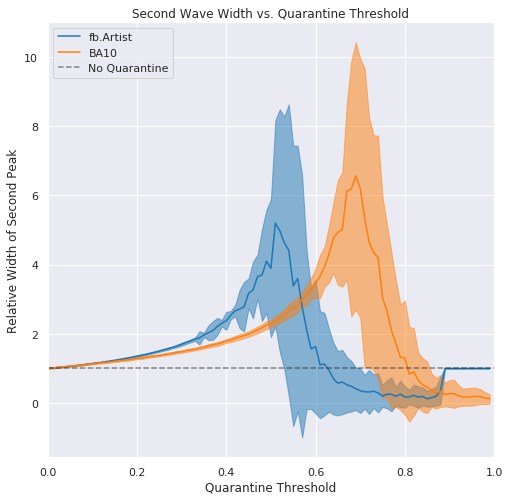} }}%
    \caption{We plot the probability of a second wave occurring for the BA network and GEMSEC Facebook Artist network (left). Observe that there is a sharp dropoff as the quarantine threshold approaches the optimal. On the right, we plot the FWHM width of the second peak relative to the width of the infection curve in the no-quarantine setting. Even when suboptimal quarantines are enacted, the second wave becomes wider. }%
    \label{fig:second-wave-stats}%
    \vspace{-1em}
\end{figure}

If a quarantine is enacted before herd immunity is reached, there will be a second "wave" of the epidemic upon reinfection. In Figure \ref{fig:second-wave-stats} (left), we plot the probability that second wave occurs at all, versus the quarantine threshold. We define a second wave occurring as the event that at least $5\%$ of remaining nodes after a quarantine become Infected. Notice that there is a sharp threshold for which a second wave becomes impossible as the quarantine proceeds, and after the optimal quarantine threshold, no second wave occurs at all, which means this quarantine threshold guarantees herd immunity. In Figure \ref{fig:second-wave-stats} (right), we plot the width of the second wave relative to the width of the no-quarantine setting. Width is calculated as the full-width at half-maximum. We observe that the second wave becomes wider as we increase the quarantine threshold, indicating that it takes longer for the infection to propagate through the population, even if the quarantine is enacted suboptimally.

\begin{figure}%
    \centering
    \subfloat{{\includegraphics[width=.45\textwidth ]{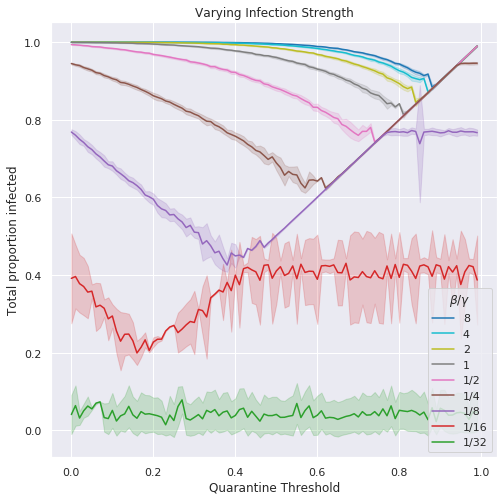} }}%
        \qquad
    \subfloat{{\includegraphics[width=.45\textwidth]{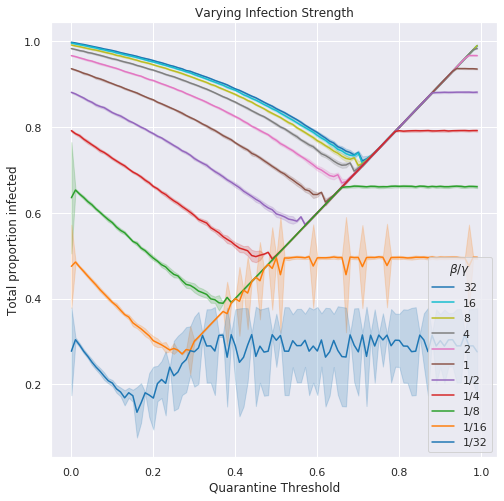} }}%
    \caption{We ablate across the infection paramaters $\beta/\gamma$ for the BA network (left) and GEMSEC Facebook Artist network (right) by plotting the total proportion of population Infected versus quarantine threshold. The qualitative properties do not change as the infection parameters are varied, though weaker infections have an optimal quarantine threshold that is earlier than a stronger infection.}%
    \label{fig:rsweep}%
\end{figure}

\paragraph{Weaker and Stronger infections} 
Until now, we have rather arbitrarily selected the ratio of infection to recovery parameters, $\frac{\beta}{\gamma}$ to be 0.5, such that the entire population becomes Infected if no quarantine is enacted. To clarify that the empirical phenomena we have discussed are not particular to this setting of infection, and to describe the effect of a stronger and weaker infection, we have run all of our simulations under the same setting where the infection parameter is varied between $\{\frac{1}{32}, \frac{1}{16}, \frac{1}{8}, \frac{1}{4}, \frac{1}{2}, 1, 2, 4, 8, 16, 32\}$. In Figure \ref{fig:rsweep} we plot the total number of Infected nodes for these varying ratios. We note that for exceptionally weak epidemics, the data is very noisy and essentially flat, indicating that epidemics are very unlikely to propagate at all. Next we observe that as an infection becomes stronger, in the sense that $\frac{\beta}{\gamma}$ is larger, the optimal quarantine threshold is higher and the quarantine is less effective. Indeed, for comparatively stronger epidemics, the trough in each V-shaped curve is shallower and further to the right. We posit that this is because the chance of the infection propagating along any single edge is much greater for strong epidemics and thus everyone becomes Infected even networks with extremely light-tailed degree distributions. Hence, even when the high-degree nodes quickly become eradicated early during the course of the epidemic, the remaining susceptible subnetwork will allow every node to become Infected, regardless of where the quarantine is applied. 

\subsubsection{Clustering and Nodes} Our theoretical results have been established in the configuration model, when networks are infinite and the clustering coefficient is vanishing. By varying the size of the network and the clustering coefficient, we prove those assumption have negligible impact on the final fraction of Removed nodes. See Appendix for exact plots.

\subsection{Multiple Quarantines}

\begin{figure}%
    \centering
    \subfloat{{\includegraphics[width=.45\textwidth ]{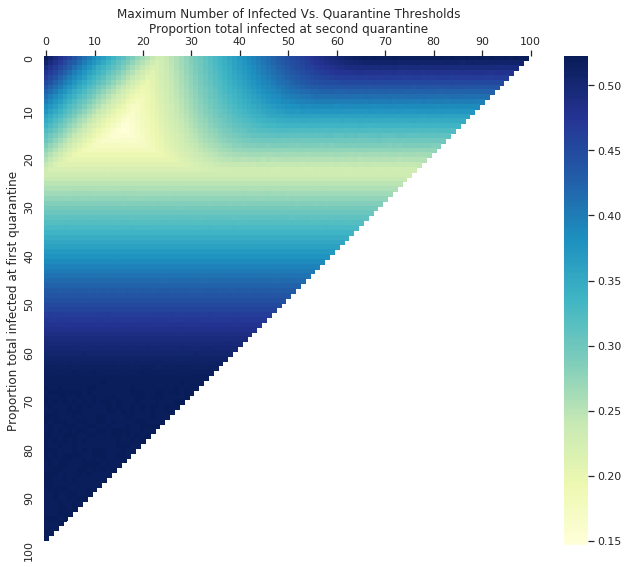} }}%
        \qquad
    \subfloat{{\includegraphics[width=.45\textwidth]{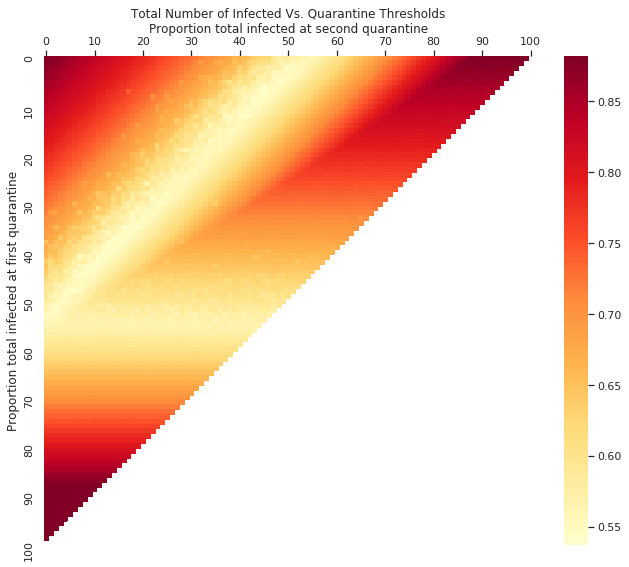} }}%
    \caption{Effect of two quarantines on the GEMSEC Facebook Artist network. The left plot displays maximum number of Infected nodes, and the right plot displays the total number of Infected nodes. The vertical axis corresponds to the proportion of the population who has been Infected at the time of first quarantine. The horizontal axis corresponds to second quarantine threshold, in terms of the proportion of the (total) population who has been Infected since the first quarantine. Observe that there is an optimal strategy in the two-quarantine setting that does better than the single-quarantine setting for the maximum Infected nodes, whereas it in general impossible to do better than the single quarantine setting for minimizing total number of Infected nodes.}%
    \label{fig:fb-heatmaps}%
    \vspace{-0.5em}
\end{figure}

\subsubsection{Two Quarantines}

We examine the effects of performing two quarantines.
We plot these results for the GEMSEC Facebook Artist graphs in Figure \ref{fig:fb-heatmaps} (see Appendix for similar result on Barabasi-Albert graphs). A two-quarantine simulation is run in a similar fashion to a single-quarantine simulation, where the second quarantine threshold refers to the proportion of the original network that becomes Infected between the first and second quarantine. The heatmaps in Figure \ref{fig:fb-heatmaps} display the first quarantine threshold on the vertical axis and the second quarantine threshold on the horizontal axis, thereby recovering the V-plots of Figure \ref{fig:single-v} in either the left column or the top row of the heatmaps. From there, we can conclude that having two quarantines can reduce the maximum number of Infected nodes, but not the Total number of Infected nodes. On one hand, this is disappointing, as we cannot reduce the final number of Removed nodes $R(u)$ by using more quarantines. On the other hand, this means that if we need to declare a quarantine earlier than at the optimal time, say because of a constraint on the maximum number of simultaneously Infected nodes (e.g. hospital beds), we can recover all the benefits of a single well-timed quarantine with two well-timed quarantines. We now extend this observation to more than two quarantines.

% Examining these heatmaps we observe two distinct qualitative phenomenon. In the 'maximum Infected heatmaps' on the left, there is indeed a benefit to enacting multiple quarantines. For the BA graph, the optimal two-quarantine setting can reduce the maximum number of Infected nodes by 34.4\% in the BA network and 36.1\% in the Facebook network. Pictorially this is demonstrated by the lightest color appearing strictly in the interior of the plot, with a triangular phase diagram. On the other hand, a second quarantine does not greatly improve the situation with respect to the total number of Infected nodes. Indeed, the optimal two-quarantine setting can reduce the total number of Infected nodes by 1.4\% in the BA network and 3.7\% in the Facebook network. On the heatmaps, this can be observed by the diagonal light-colored band that linearly connects the optimal single-quarantine thresholds in the left-column and top-row. Thus we can conclude that an additional quarantine allows for greater control over the maximum number of Infected nodes, but not much control over the total number of Infected nodes.  

\subsubsection{More than two quarantines}

\begin{figure}%
    \centering
    \subfloat{{\includegraphics[width=.6\textwidth]{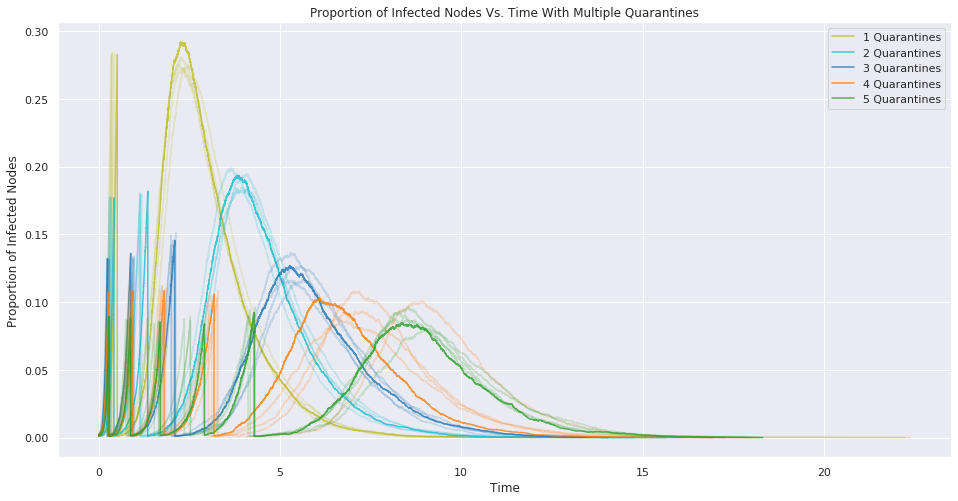} }}%
    \qquad 
    \subfloat{{\includegraphics[width=.3\textwidth]{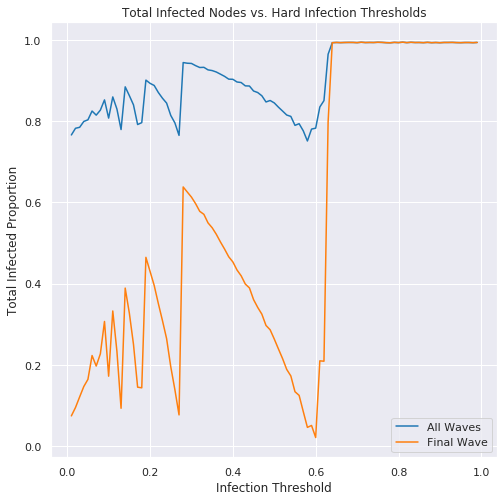} }}
    \caption{In the left plot, we map plot the number of Infected nodes versus time when multiple quarantines are allowed. The quarantine strategy is chosen to keep peak heights roughly equal. Notice that the final peaks get wider and there is diminishing returns in peak reduction as quarantines are increased. On the right, we plot the total number of Infected nodes when a hard infection threshold is enacted. Notice that the final number of nodes Infected is greatly dependent upon the number of nodes Infected by the final wave, and the optimal situation is to have an impotent final wave.}%
    \label{fig:vanilla-multiq}%
\end{figure}

Finally we turn our attention to the case when more than two quarantines are applied. In Figure \ref{fig:vanilla-multiq} (left) we plot the proportion of Infected nodes versus time for the BA10 graph under various scenarios with multiple quarantines. The quarantine thresholds are chosen to roughly equalize the peak heights. Naturally we notice that the peak heights decay as we increase the number of quarantines we're allowed, however the improvement in peak height has diminishing returns. We also notice that the latter peaks are wider than the initial peaks, echoing our observation in the single-quarantine setting. 

In Figure \ref{fig:vanilla-multiq} (right), instead of declaring a quarantine based on the fraction of nodes not in the Susceptible state, we declare it when the number of nodes in the Infected state reaches a certain threshold, and then restart a new infection after each quarantine. We then plot the final number of Removed nodes after these multiple quarantines. Notice that this strategy does not need any knowledge of the graph (not even the degree sequence). A multiple value of the threshold allow us to recover the result of a well-timed single quarantine. However, if the threshold is poorly chosen, the final number of Removed nodes can be multiple times greater than the optimal achievable. The higher the threshold, the worse this strategy can behave. Nevertheless, if quarantines are declared early (when a small fraction of the population is Infected), we are always within a factor 2 of the optimal number of nodes in the Removed state. Multiple quarantines can therefore be used as a targeted immunization strategy even without any knowledge of the contact graph, as long as they are declared early.
% One concern during a global pandemic is that the healthcare system could become overwhelmed by a massive number of simultaneously Infected patients. To remedy this, one could imagine a quarantine strategy that enforces a hard limit upon the number of simultaneously Infected nodes, where a quarantine is enacted every time the number of Infected nodes reaches the threshold. In Figure \ref{fig:vanilla-multiq} (right), we examine the efficacy of such a quarantine strategy. The horizontal axis displays the hard-limit of simultaneously Infected nodes, and the vertical axis displays the total number of Infected nodes at termination. We plot both the total proportion of Infected nodes, and the proportion of nodes that are Infected in the final wave, after all quarantines are performed. The final wave plot indicates that much of the variation in the total proportion of Recovered nodes is controlled by the last wave, which by definition must have a peak height less than the infection threshold. The optimal single quarantine scenario is represented by the right-most trough on the plot, and we notice that even when many quarantines are performed the total proportion of Infected nodes cannot be improved over the optimal single-quarantine scenario. Further, the alignment of the troughs demonstrates that the optimal strategy to minimize the total proportion of Infected nodes is to minimize the number of nodes Infected by the second wave.

\newpage
\bibliographystyle{ACM-Reference-Format}
\bibliography{all,graph}

\newpage
\appendix
\section{Appendix}
\subsection{$d$-regular graphs cannot be immunized through quarantines}
\begin{proposition}
For random $d$-regular graphs in the configuration model, i.e. with $p_d = 1$ and for $k\neq d, p_k = 0$, it is possible to achieve herd immunity after one quarantine for $d>2$. 
\proof 
For  random $d$-regular graphs in the configuration model, we have:
\begin{align*}
    g_0(x) &= x^d.
\end{align*}
Using Claim \ref{cl:generalQuarantine}, we need to declare a quarantine when the fraction of nodes of degree 1 $u$ is such that:
\begin{align*}
   0 &\geq u^2 g_0''(u) - u g_0'(u) \\
   &= u^2 d (d-1) u^{d-2} - u d u^{d-2} \\
   &= u^d d (d-2).
\end{align*}
This inequality cannot be satisfied for $k>2$, so quarantining is not a valid immunization strategy.
\\\qed
\end{proposition}
\subsection{BA graphs}
We first compare groupwise survival rates for BA graphs and a social network, then the proportion of simultaneously Infected nodes versus time. The behavior is similar for both networks.
\begin{figure}[H]
    \centering
    \subfloat{{\includegraphics[width=.45\textwidth ]{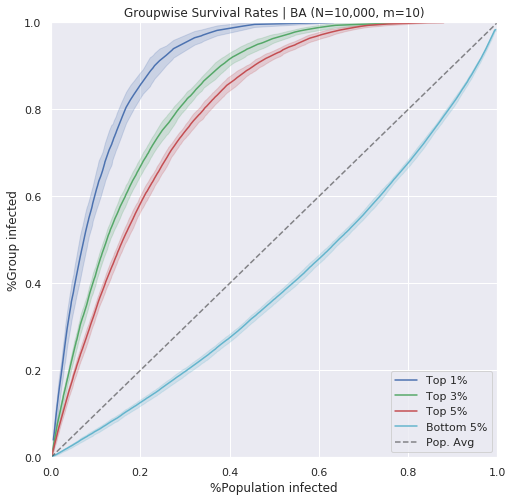} }}%
        \qquad
    \subfloat{{\includegraphics[width=.45\textwidth]{figures/groupwise_gemsec_fb_artist.png} }}%
    \caption{Groupwise survival rates for a BA graph (left) and a social network (right). Observe that the higher degree nodes become Infected much more quickly than the population average.}%
    \label{fig:groupwise-survival}%
\end{figure}

\begin{figure}[H]
    \centering
    \subfloat{{\includegraphics[width=.45\textwidth ]{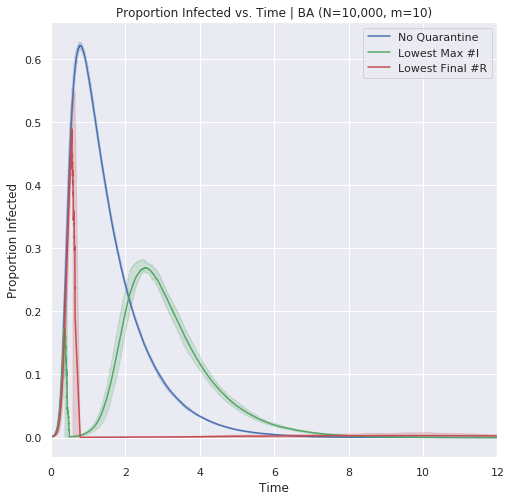} }}%
        \qquad
    \subfloat{{\includegraphics[width=.45\textwidth]{figures/vanilla_I_fbartist.png} }}%
    \caption{Proportion of simultaneously Infected population versus time for a BA network with parameter $m=10$ (left) and the GEMSEC Facebook Artist network (right). The Blue curves indicate the no quarantine setting; the green curves minimize the peak height, whereas the red curves minimize the total number of Infected nodes.}%
    \label{fig:vanilla-singleQ}%
\end{figure}

The behavior of a double quarantine is also very similar to the behavior on the GEMSEC Facebook Artist network.
\begin{figure}[H]
    \centering
    \subfloat{{\includegraphics[width=.45\textwidth ]{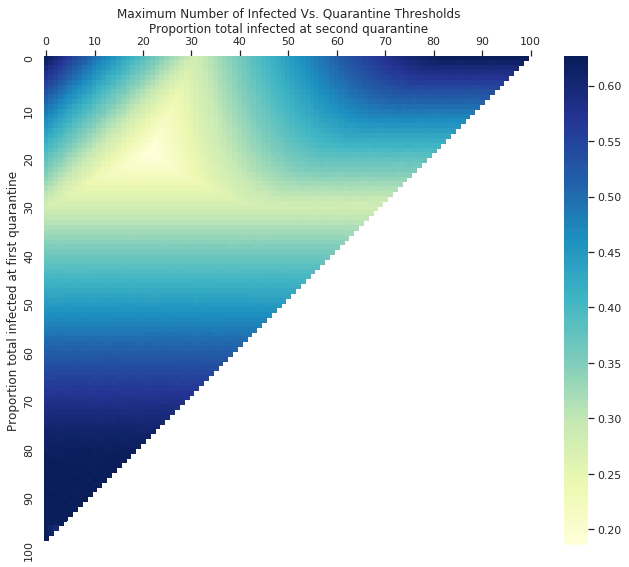} }}%
        \qquad
    \subfloat{{\includegraphics[width=.45\textwidth]{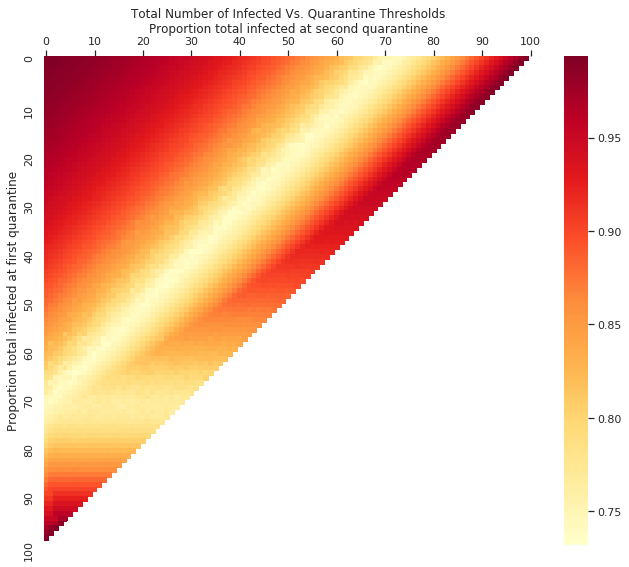} }}%
    \caption{Effect of two quarantines on the BA graph. Effects are similar to the GEMSEC Facebook Artist graphs.}%
    \label{fig:ba-heatmaps}%
\end{figure}

\subsection{Relaxing theoretical assumptions}
\begin{figure}[H]
    \centering
    \subfloat{{\includegraphics[width=.45\textwidth ]{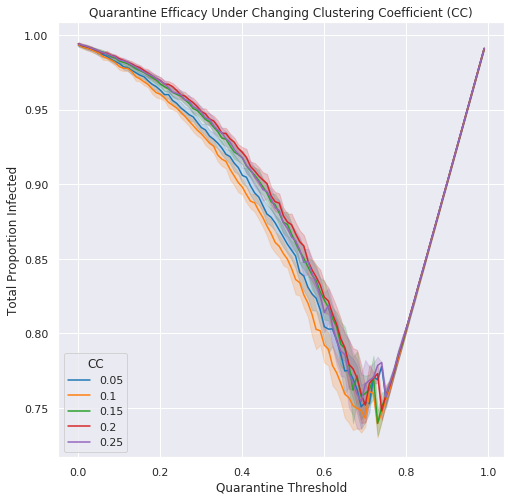} }}%
        \qquad
    \subfloat{{\includegraphics[width=.45\textwidth]{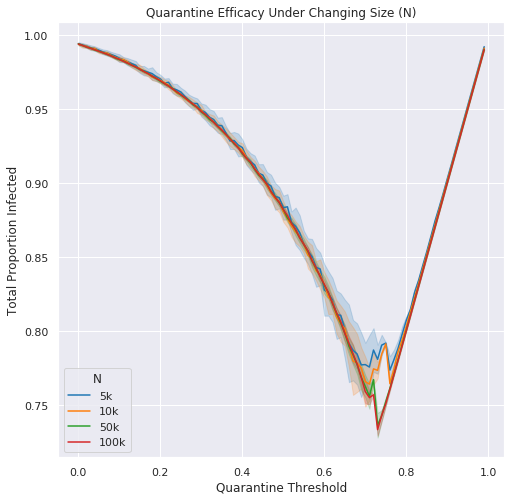} }}%
    \caption{The effect of quarantines does not depend upon clustering coefficient or size of the network. }%
    \label{fig:cluster-series}%
\end{figure}

Even though our theoretical results have been established in for infinite graphs with vanishing clustering coefficient, our empirical results hold for finite networks, and we demonstrate empirically that modifying the clustering coefficient doesn't have a noticeable effect on quarantine efficacy. In Figure \ref{fig:cluster-series} (left), we plot the total number of Infected nodes versus quarantine thresholds for a series of PLC networks, where we alter only the parameter $p$ which controls the clustering coefficient. In Figure \ref{fig:cluster-series} (right), we consider a series of BA networks with the same parameter $m$, but differing sizes. In both cases we notice that the differences across either series are negligible, indicating that the qualitative results we've empirically demonstrated do not depend upon the clustering coefficient or size of the network.

\end{document}